\documentclass[journal=jacsat,manuscript=article]{achemso}
\usepackage[version=3]{mhchem} 
\usepackage{soul}
\usepackage{threeparttable}

\author{Kerem Anar}
\affiliation{Graduate School of Sciences and Engineering (GSSE), Ko\c{c} University, Rumelifeneri Yolu, Sar{\i}yer 34450, Istanbul, T\"urkiye}

\author{Berna Akgenc Hanedar}
\affiliation{Department of Physics, K{\i}rklareli University, 39100, K{\i}rklareli, T\"urkiye}
\alsoaffiliation{Department of Physics, Ko\c{c} University, Rumelifeneri Yolu, Sar{\i}yer 34450, Istanbul, T\"urkiye}

\author{Roya Kavkhani}
\affiliation{Graduate School of Sciences and Engineering (GSSE), Ko\c{c} University, Rumelifeneri Yolu, Sar{\i}yer 34450, Istanbul, T\"urkiye}

\author{Mehmet Cengiz Onba\c{s}l{\i}}
\affiliation{Department of Electrical \& Electronics Engineering, Ko\c{c} University, Rumelifeneri Yolu, Sar{\i}yer 34450, Istanbul, T\"urkiye}
\alsoaffiliation{Department of Physics, Ko\c{c} University, Rumelifeneri Yolu, Sar{\i}yer 34450, Istanbul, T\"urkiye}
\email{monbasli@ku.edu.tr}

\email{monbasli@ku.edu.tr}

\title{Predictive Design of Defect States in Hexagonal Boron Nitride for Telecommunication-Band Quantum Emission}

\abbreviations{IR,NMR,UV}
\keywords{American Chemical Society, \LaTeX}

\begin{document}







\begin{abstract}
Defect-based single-photon emitters (SPEs) in hexagonal boron nitride (\textit{h}-BN) are promising platforms for integrated quantum photonics; however, the absence of identified emitters operating at telecom wavelengths remains a critical limitation for fiber-based quantum communication. Here, we investigate previously unexplored carbon- and silicon-based point defects in monolayer \textit{h}-BN as SPE candidates using hybrid density functional theory, constrained excited-state relaxations, and a generating-function approach to photoluminescence. We compute zero-phonon-line (ZPL) energies, radiative lifetimes, Huang–Rhys (HR) factors, and photoluminescence lineshapes to screen optical performance. All defects are thermodynamically stable with negative formation energies, and five candidates exhibit moderate electron–phonon coupling (HR $<$ 5), indicating narrow emission linewidths. These emitters span a broad spectral range from the visible to the telecom regime, including near-infrared C-based centers and, most notably, the Si$_{2\mathrm{B}}$V$_{\mathrm{N}}$ defect, which is identified as the first point defect in monolayer \textit{h}-BN predicted to support single-photon emission in the telecom C band (1554 nm). Vacancy-containing complexes possess spin-1/2 ground states, enabling spin–photon interfaces compatible with integrated photonic and cavity-based platforms. The combined analysis of ZPL energies, electron–phonon coupling, and radiative lifetimes provides concrete targets for experimental realization of spin-active single-photon emitters in \textit{h}-BN from the visible to telecom wavelengths.
\end{abstract}

\section{Introduction}

Single-photon emitters (SPEs) are essential building blocks for quantum communication, distributed quantum networks, and quantum photonic technologies.\cite{gao2023atomically,awschalom2018quantum,pompili2021realization} 
In particular, practical long-distance quantum networks rely on low-loss optical fibers, which operate most efficiently within the telecommunication wavelength bands (1.3--1.55~$\mu$m). A key challenge in solid-state quantum photonics is therefore the realization of stable, room-temperature SPEs whose emission lies within these telecom windows. 
Several solid-state platforms have demonstrated single-photon emission, including color centers in diamond and silicon carbide, as well as localized emitters in two-dimensional materials. However, most of these systems operate outside the telecom bands or require cryogenic conditions, limiting their applicability for fiber-based quantum communication.\cite{kurtsiefer2000stable,castelletto2014silicon,he2015single,srivastava2015optically,koperski2015single}

Hexagonal boron nitride (\textit{h}-BN) has emerged as a particularly attractive host for solid-state quantum emission because it combines a wide bandgap with a two-dimensional, van der Waals architecture that is naturally compatible with heterogeneous integration.\cite{caldwell2019photonics,ogawa2023hexagonal,kim2020hybrid} 
Following the initial observation of quantum emission from \textit{h}-BN,\cite{tran2016quantum,tran2016robust} a broad diversity of bright and photostable emitters has been reported, spanning ultraviolet to visible wavelengths.\cite{bourrellier2016bright,hayee2020revealing} 
Subsequent experiments have clarified key aspects of their optical performance, including narrow and, in some cases, Fourier-transform-limited lines,\cite{dietrich2018observation} resonant excitation pathways,\cite{tran2018resonant} and the roles of phonon dephasing and spectral diffusion.\cite{white2021phonon,akbari2021temperature} 
Considerable effort has also been devoted to identifying microscopic defect origins and to deterministic defect engineering. Carbon-related emitters have been implicated in multiple studies,\cite{mendelson2021identifying,zhong2024carbon} and site-controlled or structured-defect approaches have enabled improved reproducibility.\cite{gale2022site,tang2025structured,singla2024probing,lamprecht2025single} 
Beyond materials identification, the atomically thin nature of \textit{h}-BN enables efficient coupling to photonic environments, including cavity-enhanced emission and fiber-cavity architectures,\cite{parto2022cavity,haussler2021tunable,scheuer2023enhanced} providing a promising route toward integrated quantum photonic devices.

Despite this rapid experimental progress, two bottlenecks continue to limit the practical deployment of \textit{h}-BN emitters as engineered quantum light sources: (i) the microscopic identity of many emitters remains uncertain, and (ii) predicting device-relevant optical figures of merit from first principles, including the zero-phonon line (ZPL) energy, Debye--Waller factor, electron--phonon coupling strength, and radiative lifetime, remains nontrivial.\cite{sajid2020single,ping2021computational,jin2021photoluminescence} 
Bridging experiments to specific atomic structures therefore requires a quantitatively reliable theoretical description of optical excitations and vibronic line shapes associated with defect centers.\cite{onida2002electronic,alkauskas2012first,alkauskas2014first}  In this work, we address these challenges through a systematic first-principles investigation of candidate point defects in monolayer \textit{h}-BN, combining (i) ground-state electronic structure calculations to resolve defect-induced levels and spin configurations, with (ii) a configuration-coordinate framework for defect luminescence to compute photoluminescence line shapes and mode-resolved electron--phonon coupling.\cite{huang1950theory,kubo1955application,alkauskas2012first,alkauskas2014first} 
This workflow enables direct comparison between theory and experiment via experimentally accessible observables, such as ZPL positions and phonon sideband structures, and provides a practical basis for defect screening toward quantum photonic applications, as developed and applied below.\cite{ping2021computational,kim2020hybrid}

Motivated by recent experimental and theoretical advances, we focus on carbon- and silicon-related point defects in monolayer \textit{h}-BN as representative defect families with contrasting physical characteristics. Carbon impurities have been repeatedly implicated as the microscopic origin of visible and near-infrared emission in \textit{h}-BN, supported by correlated optical spectroscopy, electron microscopy, and first-principles modeling.\cite{hayee2020revealing,mendelson2021identifying,zhong2024carbon} In particular, carbon dimers and vacancy-associated carbon complexes have been proposed to account for several experimentally observed emission bands and polarization properties.\cite{mackoit2019carbon,jara2021first,golami2022b,ganyecz2024first} In parallel, silicon incorporation into \textit{h}-BN has been demonstrated experimentally and predicted theoretically to yield thermodynamically stable defect configurations with distinct electronic and magnetic properties.\cite{huang2012defect,weston2018native,ajaybu2021first} Owing to the larger atomic radius of Si compared to C, silicon-related defects are expected to host more spatially extended defect-localized wavefunctions, which can reduce exchange splitting, modify electron--phonon coupling, and potentially enable emission at longer wavelengths.\cite{ivady2018first} Despite these indications, a systematic side-by-side comparison of carbon- and silicon-based point defects using a consistent first-principles framework, particularly with respect to vibronic line shapes, Debye--Waller factors, and radiative lifetimes, has remained largely unexplored.

First-principles calculations have played a central role in assigning experimentally observed quantum emitters in \textit{h}-BN to specific atomic defects and in predicting new candidates. However, the quantitative reliability of predicted optical properties depends sensitively on the underlying electronic-structure methodology. Semi-local functionals such as PBE substantially underestimate the \textit{h}-BN band gap, while hybrid functionals such as HSE06 yield values closer to experiment, leading to systematic shifts in predicted zero-phonon-line (ZPL) energies. A wide range of defect configurations has been investigated, including vacancies, substitutional impurities, and impurity complexes. Nevertheless, reliable defect identification remains challenging because multiple defects can exhibit similar optical signatures within computational uncertainty. As a result, agreement with a single observable is insufficient for unambiguous assignment. As emphasized by Cholsuk \textit{et al.}, robust identification requires simultaneous agreement across multiple experimentally accessible properties, including ZPL energy, vibronic line shape, electron--phonon coupling strength, transition dipole characteristics, and radiative lifetime. Notably, despite extensive prior theoretical work, no point defect in monolayer \textit{h}-BN has been predicted to support single-photon emission in the telecommunication bands, underscoring a persistent gap between first-principles predictions and application-relevant spectral regimes.

In this work, we present a comprehensive first-principles investigation of carbon- and silicon-based point defects in monolayer \textit{h}-BN aimed at identifying and benchmarking candidate single-photon emitters across the visible, near-infrared, and telecommunication spectral ranges. Using density functional theory combined with a configuration-coordinate formalism for defect luminescence, we systematically evaluate defect formation energies, electronic and spin configurations, zero-phonon line (ZPL) energies, electron--phonon coupling strengths, Debye--Waller factors, and radiative lifetimes.\cite{alkauskas2014first,tawfik2022pyphotonics} By applying consistent, experimentally motivated screening criteria across these observables, we identify a subset of defects that simultaneously balance optical quality, spin functionality, and thermodynamic stability. Notably, our results predict a specific silicon-related defect in monolayer \textit{h}-BN that supports single-photon emission in the telecommunication C band with moderate electron--phonon coupling, representing, to our knowledge, the first quantitatively validated point-defect candidate in this material system to meet all key criteria for fiber-compatible quantum emission. This prediction directly addresses the long-standing absence of telecom-band emitters in \textit{h}-BN identified from first principles. Together, these findings establish a quantitative framework for defect identification in two-dimensional wide-bandgap materials and provide concrete guidance for future experimental realization and photonic integration of \textit{h}-BN-based spin--photon interfaces.\cite{awschalom2018quantum,pompili2021realization}

\section{Computational Methodology}

All calculations were performed within a first-principles density functional theory (DFT) framework using the Vienna \textit{ab initio} Simulation Package (VASP)~\cite{kresse1993ab,kresse1996efficient}. Ground-state electronic structures and total energies were obtained using a plane-wave basis and periodic supercell approach, with exchange--correlation effects described by the generalized gradient approximation (GGA) with the Perdew--Burke--Ernzerhof (PBE) functional~\cite{perdew1996generalized} and projector augmented-wave (PAW) pseudopotentials~\cite{blochl1994projector}. Defective monolayer \textit{h}-BN was modeled using a \(5\times5\times1\) supercell, chosen to balance computational tractability with sufficient spatial separation between periodically repeated defects, and a vacuum spacing of approximately \(\sim 20~\text{\AA}\) out-of-plane was introduced along the out-of-plane direction to eliminate spurious interlayer and interdefect interactions. All atomic positions were fully relaxed until the total energy change between ionic steps was below \(10^{-5}\)~eV and residual stresses were below \(\le 1\,\mathrm{kbar}\), to ensure well-converged equilibrium geometries. A plane-wave kinetic energy cutoff of 500 eV and a \(3\times3\times1\) \(k\)-point mesh were used for structural optimizations. These parameters were verified to provide converged defect formation energies and local geometries.

Defects are classified according to dopant species, substitutional site, and vacancy configuration to enable a systematic comparison of their electronic and spin properties. Structures containing two substitutional dopants X = {C, Si} occupying boron or nitrogen sites are denoted X${\textsubscript{2B}}$ and X${\textsubscript{2N}}$, respectively, and are modeled in their neutral charge states. To assess the role of vacancy-induced magnetism and symmetry breaking, configurations incorporating a single neighboring vacancy are denoted X${\textsubscript{2B}}$V$\textsubscript{N}$ (two X atoms on B sites with one N vacancy) and X${\textsubscript{2N}}$V$\textsubscript{B}$ (two X atoms on N sites with one B vacancy).

Ground-state spin configurations were determined by explicitly comparing energetically relevant spin manifolds appropriate to the electron count of each defect. For vacancy-free structures with an even number of electrons, singlet (S = 0) and triplet (S = 1) states were examined, whereas vacancy-containing structures with an odd number of electrons were evaluated in the doublet (S = 1/2) and quartet (S = 3/2) manifolds. Each spin state was realized by imposing a fixed total magnetization during self-consistent field cycles and fully relaxing the ionic positions from spin-polarized initial guesses. The lowest converged total energy was taken to define the ground-state spin configuration.

Following structural optimization, thermodynamic stability was quantified by per-atom formation energy, defined as 
\begin{align}
E_{\mathrm{for}} &= \frac{E_{\mathrm{def}} - n_{B}\mu_{B} - n_{N}\mu_{N} - 2\mu_{X}}{n_{B}+n_{N}+2}
\label{eq:Efor}
\end{align}
where \(n_{B},n_{N}\) are the number of B and N atoms in the structure; \(E_{\mathrm{def}}\) is the total energy of the defective supercell and \(\mu_{B},\mu_{N},\mu_{X}\) are chemical potentials from elemental bulk reference phases via \(\mu=E_{\mathrm{bulk}}/N\).

In addition to thermodynamic considerations, the kinetic stability of each defect was assessed using finite-temperature \textit{ab initio} molecular dynamics (AIMD) simulations. These simulations provide a complementary criterion for experimental realizability by probing whether the defect structures remain intact under thermal fluctuations. AIMD calculations were performed with the NVT ensemble at 300 K, using a time step of 1 fs for a total simulation time of 5 ps. Throughout the simulations, all investigated defects preserve their local bonding configurations without reconstruction, dissociation, or migration, indicating that the predicted structures are kinetically stable at room temperature. This robustness against thermal perturbations supports the feasibility of forming these defects under realistic growth or post-processing conditions and justifies their consideration as viable single-photon-emitter candidates.

Having established the structural and kinetic stability of the defects, we next evaluated their optical properties relevant for single-photon emission. The analysis focuses on quantities that directly govern emitter performance, including zero-phonon-line (ZPL) energies, electron–phonon coupling strengths, radiative lifetimes, and photoluminescence lineshapes. Together, these metrics provide a physically transparent basis for screening defect candidates beyond emission energy alone, allowing us to assess linewidths, brightness, and suitability for integration into photonic and spin–photon platforms.

Transition dipole moments (TDMs) were evaluated at the Brillouin-zone center \((\Gamma)\) using \textsc{Vaspkit} post-processing toolkit~\cite{wang2021vaspkit}, which extracts momentum matrix elements from converged VASP wavefunctions. This analysis was used to establish optical selection rules, identify dipole-allowed defect transitions, and screen candidates for radiative recombination relevant to single-photon emission.

The TDM is a vector quantity that quantifies the strength and polarization of the light–matter interaction between an initial electronic state  \(\psi_i\) and a final state \(\psi_f\) during an optical transition. Within the electric-dipole approximation, the transition dipole moment is obtained from the momentum operator according to 

\begin{equation}
\boldsymbol{\mu} = \frac{i\hbar}{(\epsilon_f - \epsilon_i)\,m} \langle \psi_f | \hat{\mathbf{p}} | \psi_i \rangle,
\label{eq:tdm}
\end{equation}
where \(\epsilon_i\) and \(\epsilon_f\) are the Kohn–Sham eigenvalues of the initial and final states, respectively, \(m\) is the electron mass, and \(\hat{\mathbf{p}}\) is the momentum operator. 

The squared magnitude of the TDM, \(|\boldsymbol{\mu}|^2\), reported in Debye\(^2\), was used as a quantitative criterion to distinguish optically allowed from forbidden transitions. Transitions with vanishing or negligibly small \(|\boldsymbol{\mu}|^2\) were identified as dipole-forbidden and excluded from further excited-state analysis, while transitions with finite TDMs were retained for constrained excited-state relaxations and subsequent radiative lifetime calculations.

Several first-principles approaches are available for modeling excited-state properties of point defects, each involving distinct trade-offs between accuracy and computational cost. The GW approximation combined with the Bethe--Salpeter equation (GW--BSE) is the most rigorous framework, as it explicitly accounts for quasiparticle corrections and excitonic effects.\cite{onida2002electronic,winter2021photoluminescent} However, its high computational cost and the lack of routinely accessible excited-state forces make geometry relaxation within GW--BSE impractical for systematic defect screening.\cite{ping2021computational} 

Time-dependent DFT (TDDFT) offers a more affordable alternative and has been shown to reproduce excitation energies within approximately 50 meV of GW--BSE for representative \textit{h}-BN defect systems such as C\textsubscript{B}C\textsubscript{N}.\cite{winter2021photoluminescent} Nonetheless, excited-state structural relaxations within TDDFT are less straightforward than ground-state optimizations and are not yet routinely implemented for extended defect calculations.\cite{jin2021photoluminescence}

In this work, we therefore adopt the constrained-occupation DFT ($\Delta$SCF) approach, in which an electron is promoted between selected defect-localized orbitals and the atomic geometry is explicitly relaxed under the imposed occupation constraint.\cite{gali2008ab,alkauskas2014first} This method provides direct access to relaxed excited-state geometries with an accuracy comparable to TDDFT and enables reliable evaluation of reorganization energies and electron-phonon coupling parameters required for photoluminescence modeling.\cite{jin2021photoluminescence,cholsuk2023comprehensive} Accordingly, for each optically allowed transition identified from the transition dipole moment analysis, the corresponding excited state was constructed within the $\Delta$SCF framework by promoting a single electron between defect states while maintaining the occupation constraint throughout ionic relaxation ~\cite{gali2009theory}. 

Because semi-local functionals such as PBE underestimate defect-level splittings, zero-phonon-line (ZPL) energies were evaluated using single-point hybrid-functional calculations performed on the PBE-relaxed ground- and excited-state geometries, 
\begin{equation}
E_{\mathrm{ZPL}} = E^{\mathrm{HSE06}}_{\mathrm{exc}} - E^{\mathrm{HSE06}}_{\mathrm{gs}},
\label{eq:Ezpl}
\end{equation}
where \(E^{\mathrm{HSE06}}_{\mathrm{exc}}\) and \(E^{\mathrm{HSE06}}_{\mathrm{gs}}\) denote the total energies of the excited and ground states evaluated with the HSE06 hybrid functional, employing a 25\% fraction of exact exchange and a screening parameter of \(0.2~\text{\AA}^{-1}\).

To quantify electron–phonon coupling and its impact on photoluminescence, vibrational properties of the relaxed ground-state structures were computed using finite-displacement phonon calculations implemented in Phonopy.\cite{phonopy-phono3py-JPCM} Using the optimized ground- and excited-state geometries, the configuration coordinate displacement \(\Delta Q\), the Huang--Rhys (HR) factor \(S\), and the photoluminescence (PL) lineshape were obtained using the \textsc{PyPhotonics} package,~\cite{TAWFIK2022108222} which implements the established first-principles formalism developed by Alkauskas \textit{et al.}~\cite{alkauskas2014first}. 

The mass-weighted configuration coordinate displacement between ground and excited states is defined as
\begin{equation}
(\Delta Q)^2 = \sum_{\alpha} m_{\alpha} \left( \mathbf{R}_{e,\alpha} - \mathbf{R}_{g,\alpha} \right)^2,
\label{eq:deltaQ}
\end{equation}
where \(m_{\alpha}\) is the mass of atom \(\alpha\), and \(\mathbf{R}_{g,\alpha}\) and \(\mathbf{R}_{e,\alpha}\) are the equilibrium positions of atom \(\alpha\) in the ground and excited states, respectively. To resolve the contributions of individual vibrational modes, the atomic displacement vector was projected onto the phonon eigenmodes, yielding the mode-resolved configuration coordinate

\begin{equation}
q_k = \sum_{\alpha,j} \sqrt{m_{\alpha}} \left( R_{e,\alpha j} - R_{g,\alpha j} \right) \Delta r_{k,\alpha j},
\label{eq:qk}
\end{equation}
where \(\alpha\) runs over atomic species, \(j\) runs over Cartesian coordinates \((x,y,z)\), and \(\Delta r_{k,\alpha j}\) is the displacement eigenvector of atom \(\alpha\) along direction \(j\) for phonon mode \(k\).

The partial Huang--Rhys factor for each phonon mode \(k\) is then computed as~\cite{huang1950theory}
\begin{equation}
S_k = \frac{\omega_k q_k^2}{2\hbar},
\label{eq:partial_HR}
\end{equation}
where \(\omega_k\) is the frequency of mode \(k\). The total HR factor, which measures the average number of phonons emitted during the optical transition, is obtained by summing over all modes:
\begin{equation}
S = \sum_{k} S_k.
\label{eq:total_HR}
\end{equation}

The HR factor provides a quantitative measure of the overall electron–phonon coupling strength governing optical linewidth broadening.

The Debye--Waller (DW) factor, defined as \(\mathrm{DW} = \exp(-S)\), quantifies the fraction of emission occurring in the zero-phonon line and serves as a key figure of merit for coherent single-photon emission. A lower HR factor (higher DW factor) indicates weaker electron--phonon coupling and sharper spectral lines, which is desirable for quantum photonic applications.

The distribution of phonon modes coupled to the optical transition is described by the spectral function, \(S(\hbar\omega)\), also known as the partial HR function, and is defined as

\begin{equation}
S(\hbar\omega) = \sum_{k} S_k \, \delta(\hbar\omega - \hbar\omega_k),
\label{eq:spectral_function}
\end{equation}
which directly identifies vibrational modes responsible for phonon sidebands or the phonon modes that couple most strongly to the electronic transition. 

Next, the PL lineshape was obtained using the generating function formalism~\cite{kubo1955application}. The emission intensity as a function of photon energy is given by
\begin{equation}
I_{\mathrm{em}}(\hbar\omega) = C\,\omega^3 A(\hbar\omega),
\label{eq:PL_intensity}
\end{equation}
where \(C\) is a normalization constant and \(A(\hbar\omega)\) is the optical spectral function computed via Fourier transformation of the generating function \(G(t)\):
\begin{equation}
A(E_{\mathrm{ZPL}} - \hbar\omega) = \frac{1}{2\pi} \int_{-\infty}^{\infty} G(t) \exp(-i\omega t - \gamma|t|) \, dt,
\label{eq:spectral_A}
\end{equation}
with
\begin{equation}
G(t) = \exp\left[ S(t) - S(0) \right], \quad S(t) = \int_0^{\infty} S(\hbar\omega) \exp(-i\omega t) \, d(\hbar\omega),
\label{eq:generating_function}
\end{equation}
Here, \(\gamma\) accounts for intrinsic ZPL broadening beyond harmonic phonon contributions.

Finally, combining the TDM and ZPL energy results, the radiative decay rate (and lifetime) of the zero-phonon transition were evaluated within the electric-dipole approximation:
\begin{equation}
\Gamma_{\mathrm{rad}} \equiv \tau^{-1} = \frac{n\,E_{\mathrm{ZPL}}^{3}\,|\boldsymbol{\mu}|^{2}}{3\pi\,\varepsilon_{0}\,\hbar^{4}\,c^{3}},
\label{eq:lifetime}
\end{equation}
where \(n=1.85\) is the refractive index of \textit{h}-BN~\cite{vogl2018fabrication}, \(E_{\mathrm{ZPL}}\) is the zero-phonon line energy, \(|\boldsymbol{\mu}|^{2}\) is the squared magnitude of the \(\Gamma\)-point transition dipole moment, \(\varepsilon_0\) is the vacuum permittivity, \(\hbar\) is the reduced Planck constant, and \(c\) is the speed of light in vacuum.

\section{Results and Discussion}

\begin{figure*}[t]
\centering
\includegraphics{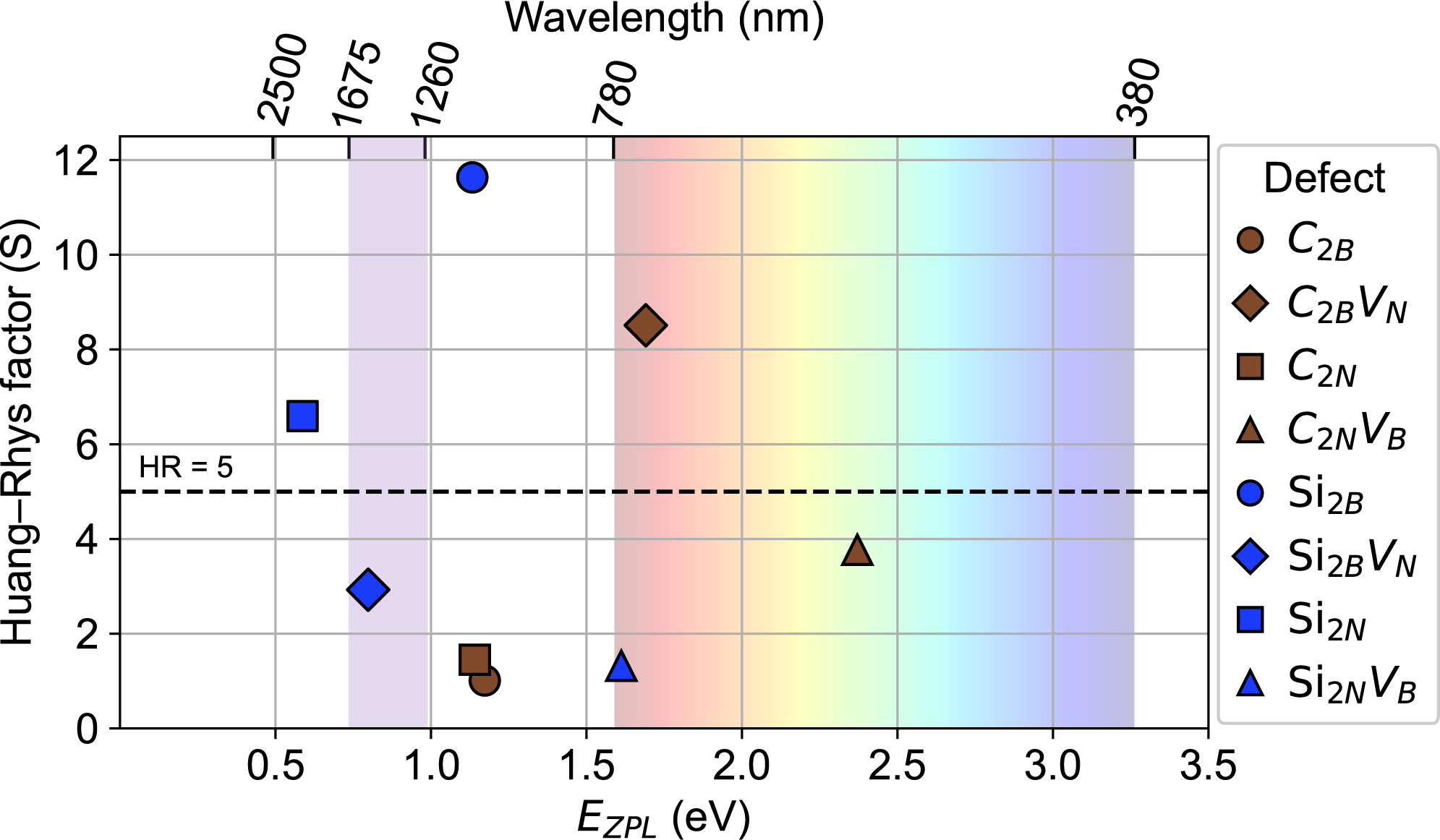}
  \caption{Huang--Rhys factor $S$ versus zero-phonon-line energy $E_{\mathrm{ZPL}}$ for all eight C- and Si-based point defects in monolayer \textit{h}-BN. The rainbow shading denotes the visible spectral range ($\sim$380--780~nm), while the purple shading marks the fiber-telecom window (O--U bands, $\sim$1260--1675~nm). The horizontal dashed line at $S=5$ indicates a boundary between intermediate and strong electron--phonon coupling. Defects below this line within the visible or telecom windows are promising candidates for narrow-linewidth single-photon emission due to their larger Debye--Waller factors.}
\label{fgr:Fig1}
\end{figure*}

\begin{table}[h]
\caption{Band gap values for pristine and defective monolayer \textit{h}-BN from experiment and theory.}
\label{tab:bandgap_comparison}
\begin{tabular}{lll}
\hline
System & Band Gap (eV) & Method \\
\hline
\multicolumn{3}{l}{\textit{Pristine \textit{h}-BN}} \\
\hline
Monolayer & 5.98\textsuperscript{\emph{d,}}\cite{watanabe2004direct} , 6.15\textsuperscript{\emph{d,}}\cite{museur2008near} , 5.5\textsuperscript{\emph{d,}}\cite{song2010large}  & Experiment \\
Monolayer & 5.96\textsuperscript{\emph{i,}}\cite{cassabois2016hexagonal} & Experiment \\
\textit{h}-BN/Cu(111) & 6.6\cite{zhang2018bandgap} & Experiment \\
Monolayer & 5.57\cite{choudhuri2018carbon} & DFT-HSE06 \\
Monolayer & 4.48\cite{choudhuri2018carbon} & DFT-PBE \\
Monolayer & 4.40\cite{ajaybu2021first} & DFT-PBE \\
Bulk & 5.95\textsuperscript{\emph{i}}, 6.47\textsuperscript{\emph{d,}}\cite{arnaud2006huge} & GW \\
Bulk & 4.02\textsuperscript{\emph{i}}, 4.46\textsuperscript{\emph{d,}}\cite{arnaud2006huge} & DFT-LDA \\
Monolayer & 4.68 (PBE), 5.77 (HSE) & This work \\
\hline
\multicolumn{3}{l}{\textit{C-doped \textit{h}-BN}} \\
\hline
C$_{\mathrm{B}}$ (3.125\%) & 1.03\cite{choudhuri2018carbon} & DFT-PBE \\
C$_{\mathrm{B}}$ (6.25\%) & 0.62\cite{choudhuri2018carbon} & DFT-PBE \\
C$_{\mathrm{B}}$ (9.375\%) & 1.84 (spin-up)\cite{choudhuri2018carbon} & DFT-HSE06 \\
C$_{\mathrm{N}}$ (3.125\%) & 4.14\cite{choudhuri2018carbon} & DFT-PBE \\
C$_{\mathrm{N}}$ (6.25\%) & 0.61\cite{choudhuri2018carbon} & DFT-PBE \\
C$_{\mathrm{N}}$ (9.375\%) & 0.76 (spin-down)\cite{choudhuri2018carbon} & DFT-HSE06 \\
C$_{2\mathrm{N}}$ & 0.14 (PBE), 1.44 (HSE) & This work \\
\hline
\multicolumn{3}{l}{\textit{Si-doped \textit{h}-BN}} \\
\hline
Si$_{\mathrm{B}}$ (6.25\%) & 3.8\cite{ajaybu2021first} & DFT-PBE \\
Si$_{\mathrm{B}}$ (12.5\%) & 4.0\cite{ajaybu2021first} & DFT-PBE \\
Si$_{2\mathrm{B}}$V$_{\mathrm{N}}$ & 1.03 (PBE), 2.02 (HSE) & This work \\
\hline
\end{tabular}

\textsuperscript{\emph{d}}direct gap;
\textsuperscript{\emph{i}}indirect gap.
\end{table}

We investigated eight C- and Si-based point defects in monolayer \textit{h}-BN as potential SPE candidates. The defect set comprises substitutional dimers occupying either the boron sublattice (nominally n-type) or the nitrogen sublattice (nominally p-type), considered both in isolation and in combination with a neighboring vacancy. To systematically screen optical performance, we used the Huang--Rhys (HR) factor $S$ as the primary metric. The HR factor quantifies the strength of electron–phonon coupling and determines the fraction of emission directed into the zero-phonon line (ZPL) versus phonon sidebands.\cite{alkauskas2014first,tawfik2022pyphotonics} 

Based on established classifications, electron--phonon coupling can be categorized as weak ($S < 1$), intermediate ($1 \le S \le 5$), or strong ($S > 5$).\cite{suyver2003temperature} Strong coupling leads to pronounced phonon sidebands and reduced spectral purity, which is detrimental for quantum photonic applications requiring indistinguishable photons.\cite{aharonovich2016solid} We therefore adopt $S < 5$ as a physically motivated threshold for identifying promising SPE candidates.
Figure~\ref{fgr:Fig1} summarizes the calculated HR factor as a function of ZPL energy for all eight defects. Five candidates satisfy this criterion and span a wide spectral range from the visible to the telecom regime. Among them, the Si$_{2\mathrm{B}}$V$_{\mathrm{N}}$ defect is particularly notable, exhibiting a ZPL within the telecom C band near 1550 nm together with moderate electron--phonon coupling. To our knowledge, this represents the first first-principles prediction of a point defect in monolayer \textit{h}-BN capable of supporting single-photon emission at telecom wavelengths. This finding positions this system as a promising platform for fiber-based quantum communication and spin-photon interfaces.\cite{awschalom2018quantum,pompili2021realization}

\begin{figure*}[h]
\centering
\includegraphics[width=16cm]{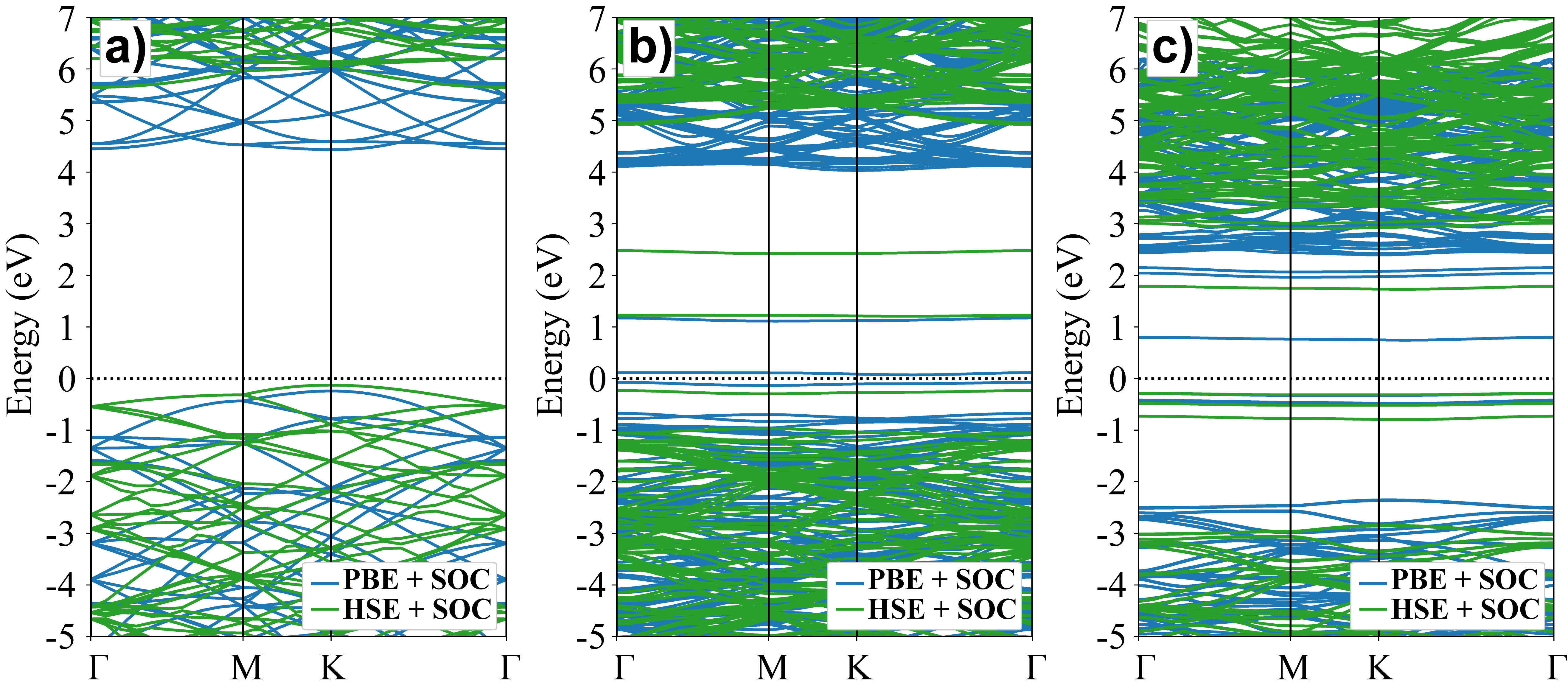}
\caption{Electronic band structures including spin--orbit coupling (SOC) (included for completeness in describing defect-state splittings), computed using PBE and HSE06 for (a) pristine monolayer \textit{h}-BN in a $5\times5\times1$ supercell, (b) the C$_{2\mathrm{N}}$ point defect, and (c) the Si$_{2\mathrm{B}}$V$_{\mathrm{N}}$ defect. Blue and green curves denote PBE+SOC and HSE06+SOC results, respectively. Bands are plotted along the high-symmetry path $\Gamma$--M--K--$\Gamma$. The horizontal dashed line indicates the Fermi energy, set to 0~eV.}
\label{fgr:Fig2}
\end{figure*}

Accurate prediction of ZPL energies is essential for identifying SPE candidates in experimentally relevant spectral windows, and the choice of exchange-correlation functional plays a critical role in achieving this accuracy. To assess the importance of hybrid functionals, we computed the electronic band structures of pristine and defective monolayer \textit{h}-BN using both PBE and HSE06 functionals with spin–orbit coupling (SOC), included for completeness in describing defect-state splittings, as shown in Figure~\ref{fgr:Fig2}. For pristine \textit{h}-BN in a 5$\times$5$\times$1 supercell, PBE yields a band gap of 4.68~eV, significantly underestimating the experimental values of 5.98--6.15~eV for monolayer \textit{h}-BN as listed in Table~\ref{tab:bandgap_comparison}.\cite{watanabe2004direct,museur2008near} In contrast, HSE06 gives a band gap of 5.77~eV, in much better agreement with experiment. This improvement is critical because ZPL energies depend sensitively on the energy difference between ground and excited states, which in turn requires an accurate description of both the host band gap and defect-induced mid-gap levels. The importance of HSE becomes even more pronounced for defective systems. For the C$_{2\mathrm{N}}$ defect shown in Figure~\ref{fgr:Fig2}, PBE predicts a band gap of only 0.14~eV, while HSE06 yields 1.44~eV—an order-of-magnitude difference. Similarly, for Si$_{2\mathrm{B}}$V$_{\mathrm{N}}$, the band gap increases from 1.03~eV with PBE to 2.02~eV with HSE06. These large corrections demonstrate that hybrid functionals are essential for reliably predicting optical transition energies in defective \textit{h}-BN and, by extension, for meaningful first-principles screening of SPE candidates. The complete set of electronic band structures and detailed band gap values for all eight defects can be found in the Supplementary Information.

All eight defects exhibit negative formation energies, indicating that they are thermodynamically favorable under appropriate growth conditions. Table~\ref{tab:properties} summarizes the calculated structural, electronic, and optical properties of all defect configurations. The formation energies span a narrow range from $-1.081$~eV for C$_{2\mathrm{B}}$ \textit{to} $-0.907$~eV for Si$_{2\mathrm{N}}$, suggesting comparable thermodynamic stability across the defect set. For C-based defects, non-vacancy configurations are slightly more favorable than the vacancy-containing counterparts, and substitution at boron sites is marginally preferred over nitrogen sites. In contrast, for Si-based defects, substitution at boron sites is clearly more favorable than at nitrogen sites. Interestingly, the introduction of a neighboring vacancy near Si dopants further lowers the formation energy, rendering Si$_{2\mathrm{B}}$V$_{\mathrm{N}}$ the most stable among the Si-based defects with a formation energy of $-1.075$~eV. Notably, this defect also emits in the telecom band, as discussed in later sections. 

Growth conditions can be tuned to favor specific doping sites. Under nitrogen-poor (boron-rich) conditions, nitrogen vacancies form more readily and substitution at nitrogen sites becomes energetically more favorable.\cite{huang2012defect,weston2018native} This suggests that the Si$_{2\mathrm{N}}$V$_{\mathrm{B}}$ configuration, although less favorable than Si$_{2\mathrm{B}}$V$_{\mathrm{N}}$ under equilibrium conditions, could still be realized by tuning the chemical potentials during epitaxy or synthesis. Importantly, all five defects identified as promising single-photon-emitter candidates based on their moderate electron–phonon coupling ($S < 5$) also possess relatively low formation energies, indicating that favorable optical properties do not come at the expense of thermodynamic stability. To assess kinetic stability, we performed finite-temperature \textit{ab initio} molecular dynamics (AIMD) simulations for all eight defects. No bond breaking, defect migration, or structural reconstruction was observed over the simulation timescale, and the total energy of each supercell remained stable throughout. These results confirm that all defects are kinetically stable at room temperature. Detailed AIMD trajectories (videos) and energy evolution plots are provided in the Supplementary Information.

\begin{table}
\caption{Summary of calculated properties for C- and Si-based point defects in monolayer \textit{h}-BN. $E_{\mathrm{for}}$: formation energy; $M$: magnetic moment; $|\mu|^2$: squared transition dipole moment; $E_{\mathrm{ZPL}}$: zero-phonon line energy; $\lambda_{\mathrm{ZPL}}$: zero-phonon line wavelength; $\tau$: radiative lifetime; $\Delta Q$: mass-weighted configuration coordinate displacement between ground and excited states; $S$: Huang--Rhys factor; DW: Debye--Waller factor.}
\label{tab:properties}
\begin{threeparttable}
\begin{tabular*}{\columnwidth}{@{\extracolsep{\fill}}lccccccccc}
\hline
Defect & $E_{\mathrm{for}}$ & $M$ & $|\mu|^2$ & $E_{\mathrm{ZPL}}$ & $\lambda_{\mathrm{ZPL}}$ & $\tau$ & $\Delta Q$ & $S$ & DW \\
 & (eV) & ($\mu_{\mathrm{B}}$) & (Debye$^2$) & (eV) & (nm) & (ns) & (amu$^{1/2}$\AA) & & \\
\hline
\textbf{C$_{2\mathrm{B}}$} & $-1.081$ & 0.00 & 11.5 & 1.173 & 1057 & 176.52  & 0.266 & 1.01 & 0.364 \\
C$_{2\mathrm{B}}$V$_{\mathrm{N}}$ & $-1.017$ & 1.00 & 37.6 & 1.691 & 733 & 18.09 & 1.234 & 8.51 & $2.0 \times 10^{-4}$ \\
\textbf{C$_{2\mathrm{N}}$} & $-1.078$ & 0.00 & 9.80 & 1.141 & 1087 & 225.63 & 0.303 & 1.45 & 0.235 \\
\textbf{C$_{2\mathrm{N}}$V$_{\mathrm{B}}$} & $-1.012$ & 1.00 & 4.64 & 2.371 & 523 & 54.07 & 0.525 & 3.77 & 0.023 \\
Si$_{2\mathrm{B}}$ & $-1.058$ & 0.00 & 9.80 & 1.133 & 1094 & 230.44 & 0.942 & 11.63 & $8.9 \times 10^{-6}$ \\
\textbf{Si$_{2\mathrm{B}}$V$_{\mathrm{N}}$} & $-1.075$ & 1.00 & 0.013 & 0.798 & 1554 & $4.93 \times 10^{5}$ & 0.475 & 2.93 & 0.053 \\
Si$_{2\mathrm{N}}$ & $-0.907$ & 0.00 & 0.55 & 0.587 & 2113 & $2.93 \times 10^{5}$ & 0.828 & 6.59 & 0.001 \\
\textbf{Si$_{2\mathrm{N}}$V$_{\mathrm{B}}$} & $-0.963$ & 1.00 & 21.1 & 1.612 & 769 & 37.19 & 0.342 & 1.32 & 0.267 \\
\hline
\end{tabular*}

\begin{tablenotes}[flushleft]
\footnotesize
\item Radiative lifetimes were evaluated within the electric-dipole approximation using a refractive index of $n=1.85$ for \textit{h}-BN.
\end{tablenotes}

\end{threeparttable}
\end{table}

\subsection{Carbon-based Point Defects}

\begin{figure}[ht]
\centering
\includegraphics[width=\columnwidth , keepaspectratio]{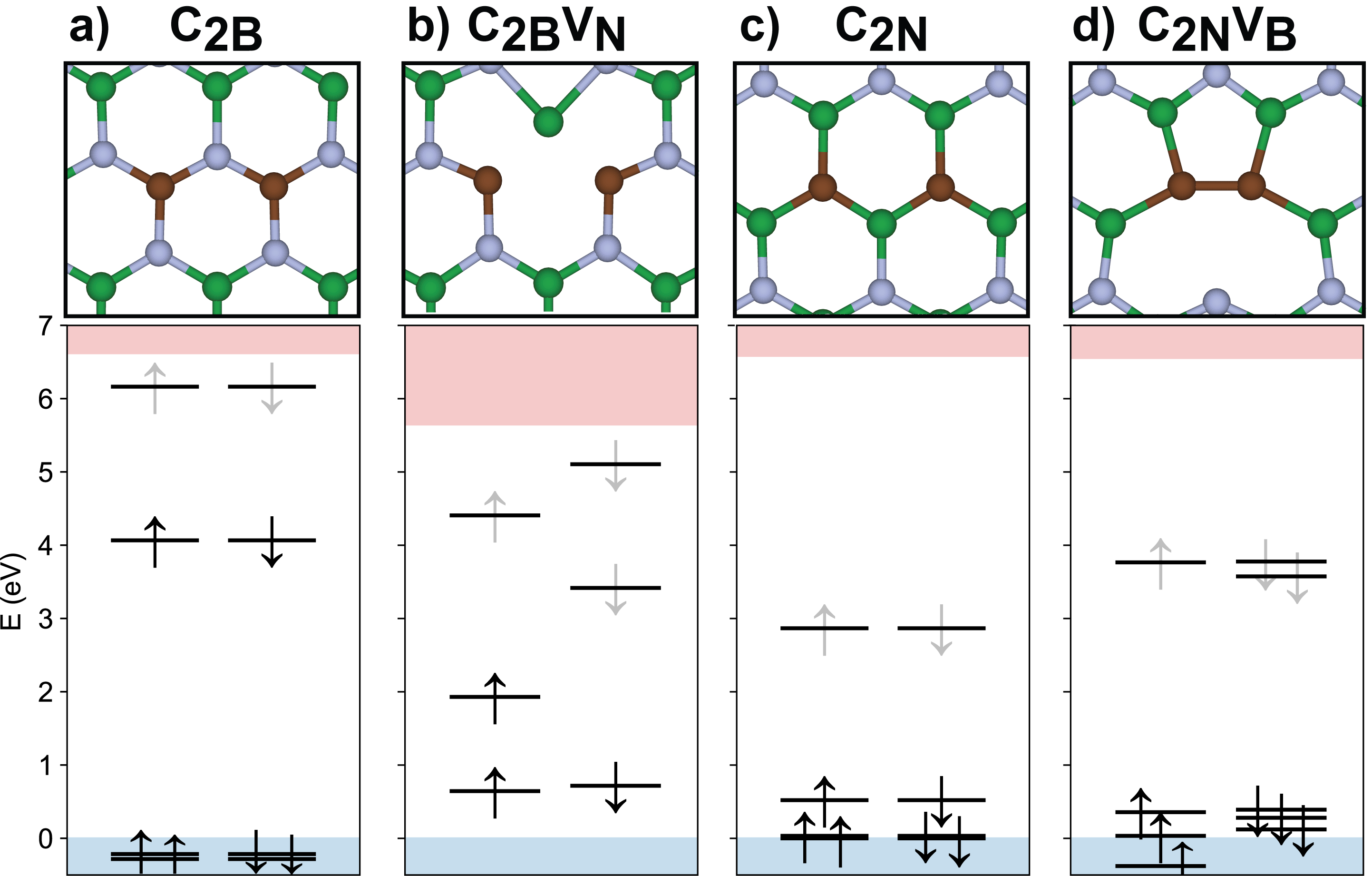}
\caption{Optimized ground-state structures (top) and corresponding single-particle Kohn--Sham defect-level diagrams (bottom) for four C-based point defects in monolayer \textit{h}-BN: (a) C$_{2\mathrm{B}}$, (b) C$_{2\mathrm{B}}$V$_{\mathrm{N}}$, (c) C$_{2\mathrm{N}}$, and (d) C$_{2\mathrm{N}}$V$_{\mathrm{B}}$. Boron, nitrogen, and carbon atoms are shown in green, gray, and brown, respectively. In the level diagrams, energies are referenced to the valence-band maximum (VBM = 0~eV). The blue and red shaded regions indicate the valence and conduction bands, respectively. Black horizontal lines denote defect-induced Kohn--Sham levels. Arrows indicate spin polarization ($\uparrow/\downarrow$); solid arrows mark occupied states, while translucent arrows indicate the corresponding unoccupied spin states. The ground states are singlet for C$_{2\mathrm{B}}$ and C$_{2\mathrm{N}}$ ($S = 0$) and doublet for C$_{2\mathrm{B}}$V$_{\mathrm{N}}$ and C$_{2\mathrm{N}}$V$_{\mathrm{B}}$ ($S = 1/2$), with total magnetic moments of 0 and 1~$\mu_{\mathrm{B}}$, respectively.}
\label{fgr:Fig3}
\end{figure}

The ground-state spin configuration of each C-based defect was determined by comparing the total energies of different spin multiplicities. For non-vacancy defects with an even number of electrons, we considered singlet ($S = 0$) and triplet ($S = 1$) states. Both C$_{2\mathrm{B}}$ and C$_{2\mathrm{N}}$ adopt a singlet ground state, described by an open-shell broken-symmetry singlet configuration to capture static correlation, which is lowest in energy. The adiabatic singlet--triplet energy gap is defined as $\Delta E_{ST}^{ad} = E_T(R_T) - E_S(R_S)$, where $E_T(R_T)$ is the triplet energy at its relaxed geometry $R_T$ and $E_S(R_S)$ is the singlet energy at its relaxed geometry $R_S$. This gap is 0.154~eV for C$_{2\mathrm{B}}$ and 0.168~eV for C$_{2\mathrm{N}}$.\cite{weber2010quantum} For the vacancy-containing defects with an odd number of electrons, we compared doublet ($S = 1/2$) and quartet ($S = 3/2$) configurations. Both C$_{2\mathrm{B}}$V$_{\mathrm{N}}$ and C$_{2\mathrm{N}}$V$_{\mathrm{B}}$ have a doublet ground state with a magnetic moment of 1~$\mu_{\mathrm{B}}$. The vertical doublet--quartet gap is defined as $\Delta E_{DQ}^{vert} = E_Q(R_D) - E_D(R_D)$, where both the quartet energy $E_Q$ and doublet energy $E_D$ are evaluated at the doublet-relaxed geometry $R_D$. This gap is 2.76~eV for C$_{2\mathrm{B}}$V$_{\mathrm{N}}$ and 1.89~eV for C$_{2\mathrm{N}}$V$_{\mathrm{B}}$. All energy gaps are much larger than the thermal energy at room temperature ($k_BT \approx 0.026$~eV), indicating thermally stable ground-state spin configurations. As shown in Figure~\ref{fgr:Fig3}, the optimized structures and single-particle defect-level diagrams clearly distinguish the nonmagnetic and spin-polarized configurations of the four C-based defects. The level diagrams in Figure~\ref{fgr:Fig3}b,d illustrate the spin-polarized electronic structure, confirming the singlet character of C$_{2\mathrm{B}}$ and C$_{2\mathrm{N}}$ and the doublet character of C$_{2\mathrm{B}}$V$_{\mathrm{N}}$ and C$_{2\mathrm{N}}$V$_{\mathrm{B}}$. The spin-1/2 ground states of the vacancy complexes make them natural candidates for spin-photon interfaces, where the electron spin can be optically addressed and coupled to emitted photons for quantum information applications.\cite{atature2018material,awschalom2018quantum}

\begin{figure}[H]
\centering
\includegraphics[width=\columnwidth , keepaspectratio]{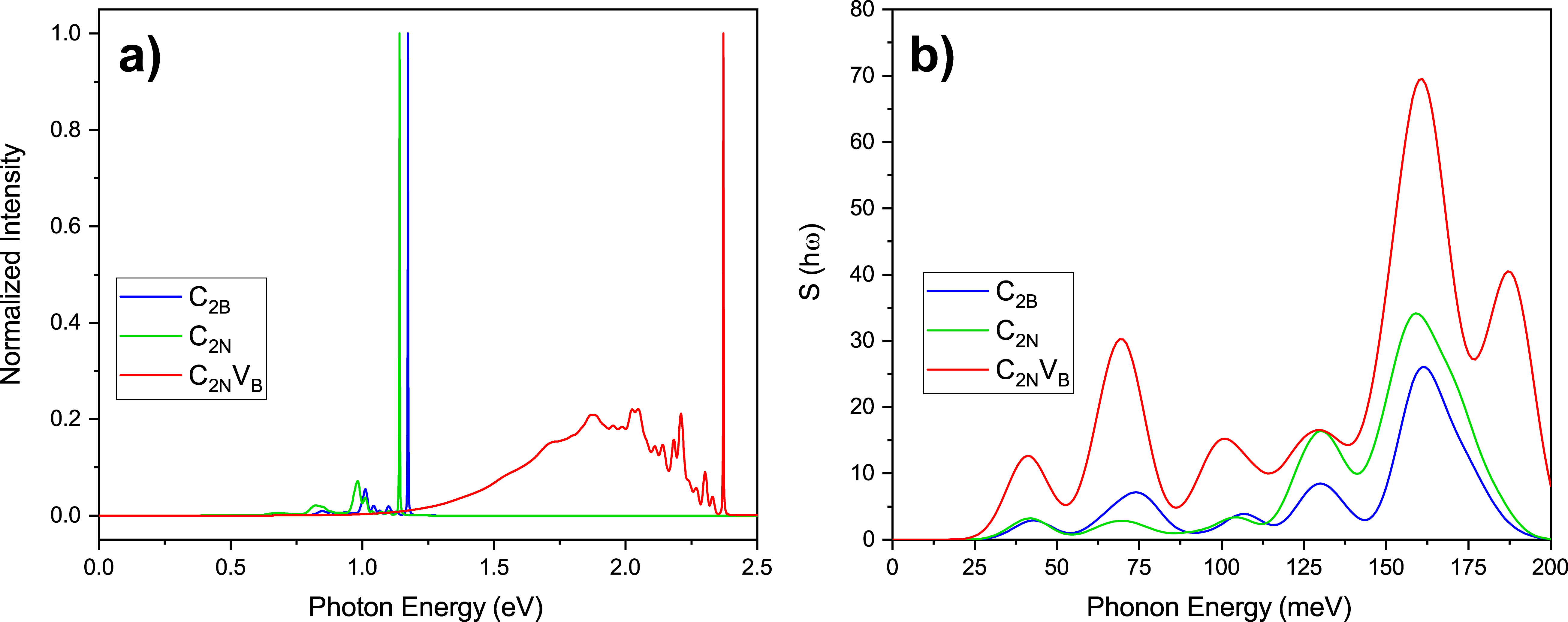}
  \caption{Calculated photoluminescence (PL) line shapes and electron--phonon coupling for C-based point defects in monolayer \textit{h}-BN. (a) Normalized emission intensity (maximum intensity set to unity) versus photon energy for C$_{2\mathrm{B}}$, C$_{2\mathrm{N}}$, and C$_{2\mathrm{N}}$V$_{\mathrm{B}}$ (excluding C$_{2\mathrm{B}}$V$_{\mathrm{N}}$ due to its strong electron--phonon coupling). (b) Partial Huang--Rhys spectra $S(\hbar\omega)$ as a function of phonon energy for the same defects, illustrating the mode-resolved coupling strengths that shape the phonon sidebands in (a).}
\label{fgr:Fig4}
\end{figure}

Among the four C-based defects, C$_{2\mathrm{B}}$, C$_{2\mathrm{N}}$, and C$_{2\mathrm{N}}$V$_{\mathrm{B}}$ satisfy the $S < 5$ criterion and are considered promising SPE candidates. The calculated optical properties are summarized in Table~\ref{tab:properties}. C$_{2\mathrm{B}}$ and C$_{2\mathrm{N}}$ emit in the near-infrared region with ZPL energies of 1.173~eV and 1.141~eV, corresponding to wavelengths of 1057~nm and 1087~nm, respectively. These wavelengths fall within the operating range of InGaAs detectors and, while outside the telecom C band, are compatible with existing fiber-based quantum optical platforms.\cite{aharonovich2016solid} Both defects exhibit weak electron--phonon coupling with HR factors close to unity, resulting in Debye--Waller factors of 0.36 for C$_{2\mathrm{B}}$ and 0.24 for C$_{2\mathrm{N}}$. These values are approximately an order of magnitude larger than the Debye--Waller factor of the NV center in diamond ($\mathrm{DW} \approx 0.03$), implying a substantially higher fraction of emission into the zero-phonon line. This means that 24--36\% of the total emission is concentrated in the ZPL for our C-based defects, compared to only 3\% for the NV center. Such high ZPL fractions are favorable for applications requiring spectral purity and photon indistinguishability. The small configuration coordinate displacements, $\Delta Q < 0.31$~amu$^{1/2}$\AA, indicate minimal structural relaxation upon optical excitation, consistent with the weak phonon coupling. The radiative lifetimes of C$_{2\mathrm{B}}$ and C$_{2\mathrm{N}}$ are 177~ns and 226~ns, respectively, comparable to those reported for other NIR-emitting defects in wide-bandgap semiconductors.\cite{castelletto2014silicon} Figure~\ref{fgr:Fig4}a shows the calculated PL lineshapes for the three promising C-based defects. C$_{2\mathrm{B}}$ and C$_{2\mathrm{N}}$ display sharp ZPL peaks with weak phonon sidebands. The partial HR spectral function in Figure~\ref{fgr:Fig4}b reveals that the dominant phonon coupling occurs around 150--170~meV. These energies correspond to optical phonon modes in \textit{h}-BN, with the peak near 170~meV resembling the E$_{2g}$ mode of pristine \textit{h}-BN.\cite{reich2005resonant,vuong2016phonon}

Introducing a vacancy significantly increases electron--phonon coupling. The C$_{2\mathrm{B}}$V$_{\mathrm{N}}$ defect exhibits a Huang--Rhys factor of $S = 8.51$, well above the adopted $S < 5$ threshold, and is therefore excluded from the set of promising single-photon-emitter candidates. The large configuration coordinate displacement, $\Delta Q = 1.23~\mathrm{amu}^{1/2}\text{\AA}$, indicates substantial structural rearrangement in the excited state, leading to broad phonon sidebands and a negligible Debye--Waller factor. In contrast, C$_{2\mathrm{N}}$V$_{\mathrm{B}}$ retains an intermediate electron--phonon coupling strength with $S = 3.77$ despite the presence of a vacancy. This defect emits in the visible range with a zero-phonon line at $E_{\mathrm{ZPL}} = 2.371~\mathrm{eV}$, corresponding to a wavelength of 523~nm, which is close to the commonly observed green emission band in \textit{h}-BN around 2.0--2.2~eV.\cite{tran2016quantum,mendelson2021identifying} Emission in this spectral range makes C$_{2\mathrm{N}}$V$_{\mathrm{B}}$ compatible with standard confocal microscopy setups employing 532~nm excitation and silicon-based single-photon detectors. Although its Debye--Waller factor ($\mathrm{DW} = 0.023$) is lower than those of the non-vacancy defects C$_{2\mathrm{B}}$ and C$_{2\mathrm{N}}$, the calculated photoluminescence lineshape in Fig.~\ref{fgr:Fig4}a shows that the zero-phonon line remains clearly distinguishable above the phonon sidebands. While the vacancy-free defects C$_{2\mathrm{B}}$ and C$_{2\mathrm{N}}$ offer superior optical properties due to their higher Debye--Waller factors, they lack an intrinsic spin degree of freedom. In contrast, C$_{2\mathrm{N}}$V$_{\mathrm{B}}$ possesses a spin-$1/2$ ground state with a magnetic moment of $1~\mu_{\mathrm{B}}$, enabling potential spin--photon coupling for quantum information applications.\cite{gottscholl2020initialization} This defect therefore represents a compromise between optical quality and spin functionality among the carbon-based candidates.

\subsection{Silicon-based Point Defects}

\begin{figure}[H]
\centering
\includegraphics[width=\columnwidth , keepaspectratio]{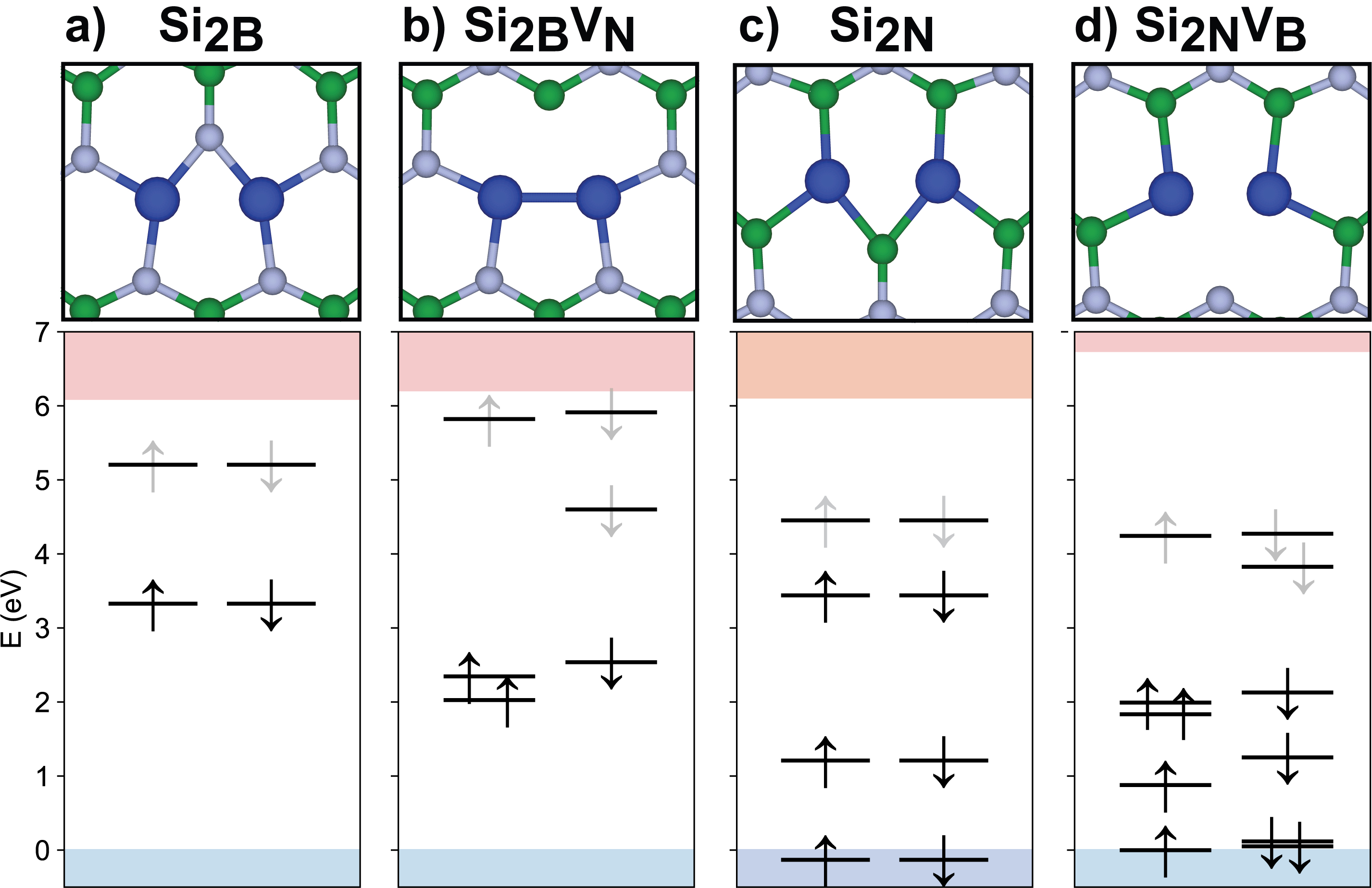}
  \caption{
  Optimized ground-state structures (top) and corresponding single-particle defect-level diagrams (bottom) for four Si-based point defects in monolayer \textit{h}-BN: (a) Si$_{2\mathrm{B}}$, (b) Si$_{2\mathrm{B}}$V$_{\mathrm{N}}$, (c) Si$_{2\mathrm{N}}$, and (d) Si$_{2\mathrm{N}}$V$_{\mathrm{B}}$. Boron, nitrogen, and silicon atoms are shown in green, gray, and blue, respectively. In the level diagrams, energies are referenced to the valence-band maximum (VBM = 0~eV). The blue and red shaded regions indicate the valence and conduction bands, respectively. Black horizontal lines denote defect-induced Kohn--Sham levels. Arrows indicate spin polarization ($\uparrow/\downarrow$); solid arrows mark occupied states, while translucent arrows indicate the corresponding unoccupied spin channel. The ground states are singlet for Si$_{2\mathrm{B}}$ and Si$_{2\mathrm{N}}$ ($S = 0$) and doublet for Si$_{2\mathrm{B}}$V$_{\mathrm{N}}$ and Si$_{2\mathrm{N}}$V$_{\mathrm{B}}$ ($S = 1/2$), with total magnetic moments of 0 and 1~$\mu_{\mathrm{B}}$, respectively.}
\label{fgr:Fig5}
\end{figure}
The spin ground states of Si-based defects were determined using the same approach described for the C-based systems. Si$_{2\mathrm{B}}$ and Si$_{2\mathrm{N}}$ have an even number of electrons and both adopt a singlet ground state with $S = 0$. Unlike the C-based analogs where the open-shell broken-symmetry singlet is clearly favored, the closed-shell and broken-symmetry configurations are nearly degenerate for the Si-based defects. This behavior can be attributed to the larger atomic radius of Si compared to C, which leads to more spatially extended defect-localized wavefunctions. Extended wavefunctions reduce the Coulomb repulsion between electrons, diminishing the energy gain from localizing electrons on separate atomic sites. The adiabatic singlet--triplet gaps are 0.074~eV for Si$_{2\mathrm{B}}$ and 0.064~eV for Si$_{2\mathrm{N}}$, smaller than those of C$_{2\mathrm{B}}$ and C$_{2\mathrm{N}}$ but still well above the thermal energy at room temperature. For the vacancy-containing defects, Si$_{2\mathrm{B}}$V$_{\mathrm{N}}$ and Si$_{2\mathrm{N}}$V$_{\mathrm{B}}$ both have a doublet ground state with $S = 1/2$ and a magnetic moment of 1~$\mu_{\mathrm{B}}$. The vertical doublet--quartet gaps are 1.87~eV for Si$_{2\mathrm{B}}$V$_{\mathrm{N}}$ and 0.98~eV for Si$_{2\mathrm{N}}$V$_{\mathrm{B}}$, smaller than those of the C-based vacancy complexes. This reduction can be attributed to weaker exchange interactions in the Si-based defects.\cite{ivady2018first} The exchange interaction depends on the spatial overlap of electron wavefunctions, and more extended wavefunctions result in smaller overlap and thus weaker spin splitting. Nevertheless, all gaps remain much larger than $k_BT \approx 0.026$~eV, ensuring thermally stable spin configurations at room temperature. Figure~\ref{fgr:Fig5} presents the optimized structures and single-particle level diagrams for the four Si-based defects. The spin-1/2 ground states of Si$_{2\mathrm{B}}$V$_{\mathrm{N}}$ and Si$_{2\mathrm{N}}$V$_{\mathrm{B}}$ open the possibility for spin-photon interface applications, similar to the C-based vacancy complexes.

\begin{figure}[H]
\centering
\includegraphics[width=\columnwidth , keepaspectratio]{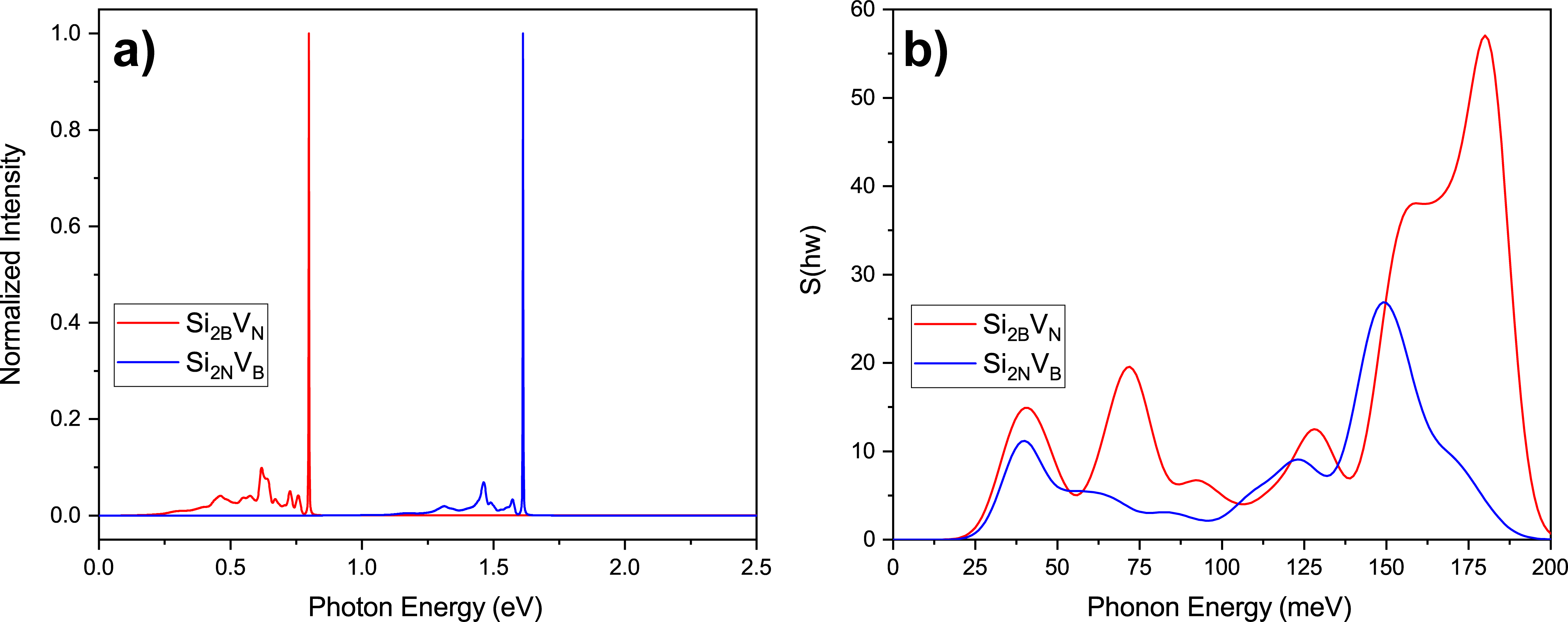}
  \caption{Calculated photoluminescence (PL) line shapes and mode-resolved electron--phonon coupling for Si-based vacancy complexes in monolayer \textit{h}-BN. (a) Normalized emission intensity versus photon energy for Si$_{2\mathrm{B}}$V$_{\mathrm{N}}$ (red) and Si$_{2\mathrm{N}}$V$_{\mathrm{B}}$ (blue). (b) Partial Huang--Rhys spectra $S(\hbar\omega)$ for the same defects, showing the phonon modes that dominate the sideband structure observed in (a).}

\label{fgr:Fig6}
\end{figure}

The photoluminescence results for the Si-based defects separate naturally into two groups. The substitutional dimers Si$_{2\mathrm{B}}$ and Si$_{2\mathrm{N}}$ have large Huang--Rhys factors and configuration coordinate displacements. Their emission is therefore dominated by broad phonon sidebands rather than by the ZPL, and they are not considered as single-photon emitter candidates in the following discussion (Table~\ref{tab:properties}). Detailed results of Si$_{2\mathrm{B}}$ and Si$_{2\mathrm{N}}$ can be found in the Supplementary Information. However, introducing a neighboring vacancy changes the picture for Si dopants. Both Si$_{2\mathrm{B}}$V$_{\mathrm{N}}$ and Si$_{2\mathrm{N}}$V$_{\mathrm{B}}$ have $S < 5$ and show well-defined ZPL peaks with reduced sidebands in Fig.~\ref{fgr:Fig6}. This trend is opposite to the C-based series, where vacancies increased the HR factor and strongly reduced the Debye--Waller factor. Within the Si-based defects, Si$_{2\mathrm{B}}$V$_{\mathrm{N}}$ is particularly notable because it is, to our knowledge, the first point defect in monolayer \textit{h}-BN predicted to support single-photon emission in the telecom band. It has a ZPL energy of 0.798~eV, which corresponds to a wavelength of 1554~nm in the telecom C band. The electron--phonon coupling is intermediate with $S = 2.93$, and the Debye--Waller factor is 0.053, so a finite fraction of the emission remains in the ZPL. The transition dipole moment is small but non-zero ($|\mu|^2 = 0.013$~Debye$^2$), which leads to a long radiative lifetime of 493~$\mu$s. This combination of long lifetime, telecom wavelength, and moderate Debye--Waller factor is similar in character to other telecom-band color centers such as the T center in silicon\cite{xiong2024computationally,bergeron2020characterization} and the chlorine vacancy in 4H-SiC,\cite{bulancea2023chlorine} which are typically integrated with cavities or waveguides to enhance their weak ZPL emission through the Purcell effect.\cite{purcell1995spontaneous} Indeed, cavity-induced Purcell enhancement is a well-established approach for emitters with long radiative lifetimes. For the T center in silicon, which has a bulk lifetime of approximately 1~$\mu$s, cavity integration has reduced the lifetime to 136~ns and improved ZPL brightness by two orders of magnitude.\cite{islam2023cavity,johnston2024cavity} Even longer-lived erbium ions (Er$^{3+}$) with radiative lifetimes exceeding several milliseconds have been transformed into practical single-photon sources through Purcell factors of 140--530.\cite{dibos2022purcell,yu2023frequency,merkel2020coherent} For defects in \textit{h}-BN, fiber-based cavities have achieved spectral enhancement factors up to 50.\cite{haussler2021tunable} Similar enhancement is expected for Si$_{2\mathrm{B}}$V$_{\mathrm{N}}$, and the two-dimensional nature of \textit{h}-BN should facilitate integration with various photonic platforms.\cite{parto2022cavity} Together with its spin-1/2 ground state and 1~$\mu_{\mathrm{B}}$ magnetic moment, these considerations make Si$_{2\mathrm{B}}$V$_{\mathrm{N}}$ a promising candidate for telecom-band spin-photon interfaces. Si$_{2\mathrm{N}}$V$_{\mathrm{B}}$ provides a complementary option. Its ZPL energy of 1.612~eV (769~nm) lies at the boundary between the visible and near-infrared regions. The HR factor is small ($S = 1.32$), and the Debye--Waller factor is relatively large at 0.267. In addition, the transition dipole moment is strong ($|\mu|^2 = 21.09$~Debye$^2$) and the radiative lifetime is only 37.19~ns. Si$_{2\mathrm{N}}$V$_{\mathrm{B}}$ is therefore a bright spin-1/2 emitter that is well suited for detection with standard silicon single-photon detectors, even though it does not operate in a low-loss fiber window.

\section{Conclusions}\label{sec:conclusion}

In this work, we have systematically investigated carbon- and silicon-based point defects in monolayer hexagonal boron nitride as candidates for defect-based single-photon emitters spanning the visible to telecom wavelength regimes. Using hybrid density functional theory combined with constrained excited-state relaxations and a generating-function formalism for photoluminescence, we evaluated key optical metrics including zero-phonon-line energies, radiative lifetimes, electron–phonon coupling strengths, and emission lineshapes. This unified framework enables a consistent comparison of candidate defects beyond ZPL energies alone.

Our results identify several thermodynamically stable defects exhibiting moderate electron–phonon coupling and favorable radiative properties, suggesting the potential for narrow-linewidth single-photon emission. Notably, the Si$_{2\mathrm{B}}$V$_{\mathrm{N}}$ defect is predicted to support emission in the telecom C band, representing, to our knowledge, the first point defect in monolayer \textit{h}-BN proposed as a telecom-wavelength single-photon emitter. In addition, vacancy-containing complexes exhibit spin-$1/2$ ground states, indicating compatibility with spin–photon interfaces in van der Waals materials.

More broadly, this study establishes a predictive, first-principles-driven pathway for identifying optically and spin-active quantum defects in two-dimensional materials across technologically relevant wavelength ranges. The identified defect candidates provide concrete targets for future experimental synthesis, optical characterization, and cavity integration, and position \textit{h}-BN as a promising platform for scalable, fiber-compatible quantum photonic technologies.

\begin{acknowledgement}

This work was carried out using the KUACC High-Performance Computing Cluster and the High-Performance Grid Computing Center (TR-Grid e-Infrastructure) at TÜBİTAK ULAKBİM. The study was supported by the European Research Council (ERC) under the Starting Grant SKYNOLIMIT (Grant No. 948063), the ERC Proof-of-Concept Grant SuperPHOTON (Grant No. 101100718), and the U.S. Air Force Office of Scientific Research, European Office of Aerospace Research and Development (Grant No. FA8655-24-1-7033).

\end{acknowledgement}

\begin{suppinfo}

The Supporting Information contains additional computational details, convergence tests, defect geometries, charge and spin densities, formation-energy data, and supplementary photoluminescence and lineshape analyses.

\end{suppinfo}

\bibliography{achemso-demo}

\end{document}